\documentclass{iopjournal}

\usepackage{amsmath}
\usepackage{graphicx}
\usepackage{amsmath,amssymb}
\usepackage{bm}
\usepackage{color}
\usepackage{tikz}
\usepackage[T1]{fontenc}
\usepackage[utf8]{inputenc}
\usepackage{booktabs}

\begin{document}

\articletype{Article type} 

\title{Thermal Phase Structure of the Attractive Fermi Hubbard Model with Imaginary Chemical Potential}

\author{Evangelos G. Filothodoros$^1$\orcid{https://orcid.org/0000-0002-5898-7288}}

\affil{$^1$Physics Department, Aristotle University of Thessaloniki, Thessaloniki, Greece}

\email{efilotho@physics.auth.gr}

\keywords{Hubbard model, imaginary chemical potential, BCS-BEC crossover}

\begin{abstract}
We study the BCS--BEC crossover of the large $N$ attractive Fermi-Hubbard model on a one-dimensional lattice using the mean field approximation in the presence of an imaginary chemical potential. We show that the crossover is governed by three parameters. The imaginary chemical potential $i\theta$, the temperature via a thermal kernel $g(\beta E_k,\beta\theta)$ and the parameter $\delta_u$ whose sign controls the weak and strong coupling regimes. At the unitarity point ($U=U_c$), we find a thermal window $\phi=\beta\theta=2\pi/3,4\pi/3$ where the gap vanishes while the fermion number $N_f$, which quantifies the balance between particle-like and hole-like excitations, has a local maximum/minimum. Inside this thermal window BCS and BEC physics are await changes in the coupling to be selected as the dominant regime. We expect that our results will unveil a better understanding of pairing correlations in lattice many-body physics.
\end{abstract}

\section{Introduction}

The BCS--BEC crossover is one of the most intriguing phenomena in quantum many-body physics \cite{Zwerger, Garg, Diener, Bauer, Watanabe}. It describes the transition from weakly coupled (BCS) superconductivity to strongly-coupled Bose-Einstein condensation (BEC). Recently, this crossover has been studied from ultracold atomic gases to high-temperature superconductors and twisted graphene multilayers \cite{Chena, Chen} but also as a microscopic model in an optical lattice \cite{Koetsier}. In order to understand these phenomena and to capture the competition between parameters like external gauge fields, pairing interactions and temperature, we use the simplified theory of the attractive Fermi-Hubbard model.

In this work, we investigate the phase crossover of the large $N$ attractive Fermi-Hubbard model on a one-dimensional lattice using mean-field approximation in the presence of an imaginary chemical potential. This imaginary chemical potential arises naturally from the canonical ensemble when one introduces a constraint on the fermion number. Although an imaginary chemical potential is not directly observable in lattice theory, it corresponds to a background $U(1)$ gauge field creating a holonomy along the thermal circle, which is analogous to Aharonov-Bohm phases \cite{Kapustin}. Surprisingly, this temporal gauge field picture connects our analysis to a variety of phenomena where fermions acquire phases from moving around closed loops.

Our main result is the discovery of a "thermal window" at the imaginary chemical potential values $\phi=\beta\theta = 2\pi/3$ and $4\pi/3$. In particular, when the system is tuned to unitarity ($U = U_c$), the superconducting gap $\Delta$ vanishes while the fermion number $N_f$, which separates particle-like from hole-like excitations, reaches a critical value. We find that the curvature of the thermodynamic potential changes sign precisely at these points, defining the boundaries between BCS and BEC regimes.

These angles are special since the condition $\cos(\beta\theta) = -1/2$ creates a pattern for fermions propagating in imaginary time. This same mathematical condition appears across a variety of physical contexts like topological band theories where staggered magnetic fluxes generate Berry phases \cite{Liu}, cold atom experiments with synthetic lattices \cite{Li} and Floquet-engineered gauge fields leading to nontrivial topological phases \cite{Wang}.

In Section 2, we introduce the large-$N$ attractive Fermi-Hubbard model with an imaginary chemical potential. In Section 3 we derive the gap equation and number equation, introducing the key parameters governing the crossover. In Section 4 we analyse the phase structure, focusing on the thermal kernel $g(\beta E_k,\beta\theta)$ and then we are able to show the emergence of the special angles. We also derive the exact values of fermion number as a function of $T/t$ ratio for small and large temperature and the value of $T/t$ for which the two expressions are equal. In Section 5 we
summarise and offer a few ideas for future work. Three Appendices contain some
technical results and useful formulae.

\section{Attractive Fermi-Hubbard Model and Large-$N$ Formulation}

We consider the spin-$\frac12$ Fermi--Hubbard model on a lattice at inverse temperature
$\beta$ \cite{Tanaka}, with a fixed total fermion number $N_f$ \cite{Kapustin} which gives the balance between particle-like and hole-like excitations and $n_{i\sigma} = c_{i\sigma}^\dagger c_{i\sigma}$, so the canonical partition function is
\begin{equation}
Z_{N_f} = \mathrm{Tr}\left[ \delta(\hat N - N_f)\, e^{-\beta H} \right],
\qquad
\hat N = \sum_{i,\sigma} n_{i\sigma}.
\end{equation}
Introducing the Fourier representation of the projector,
\begin{equation}
\delta(\hat N - N_f)
= \frac{1}{2\pi} \int_0^{2\pi} d\theta \;
e^{i\theta(\hat N - N_f)},
\end{equation}
the canonical partition function \cite{Delia, Dagotto} can be written as
\begin{equation}
Z_{N_f}
= \frac{1}{2\pi} \int_0^{2\pi} d\theta\;
e^{-i\theta N_f}\, Z(\theta),
\end{equation}
where $Z(\theta)$ is a grand-canonical partition function with an
imaginary chemical potential $\mu = i\theta$, for $\theta$ to be dimensionless.

\subsection{Pairing-channel formulation and large-$N$ effective action}
To describe the BCS--BEC crossover, the Hubbard interaction must be
decoupled in the \emph{pairing channel}, so we introduce a large-$N$ generalization by extending the fermions to $N$ flavors $\alpha=1,\dots,N$, which provides a controlled
saddle-point limit. The momentum version of the Hamiltonian is \cite{Schrieffer}
\begin{equation}
H
=
\sum_{\mathbf k,\alpha,\sigma}
\epsilon_{\mathbf k}\,
c_{\mathbf k\alpha\sigma}^\dagger
c_{\mathbf k\alpha\sigma}
-
\frac{U}{2N}
\sum_i
\sum_{\alpha,\beta}
c_{i\alpha\uparrow}^\dagger
c_{i\alpha\downarrow}^\dagger
c_{i\beta\downarrow}
c_{i\beta\uparrow},
\end{equation}
where $U>0$ is the attractive interaction. The factor $1/N$ ensures that
the interaction energy scales extensively in the large-$N$ limit.
The Euclidean action then takes the form
\begin{equation}
S=\int_0^\beta d\tau\Bigg[
\sum_{\mathbf k,\alpha,\sigma}
c_{\mathbf k\alpha\sigma}^\dagger
(\partial_\tau+\epsilon_{\mathbf k}-i\theta)
c_{\mathbf k\alpha\sigma}
-\frac{U}{2N}
\sum_{i,\alpha,\beta}
c_{i\alpha\uparrow}^\dagger
c_{i\alpha\downarrow}^\dagger
c_{i\beta\downarrow}
c_{i\beta\uparrow}
\Bigg].
\end{equation}
The quartic interaction is decoupled via a Hubbard--Stratonovich
transformation in the pairing channel \cite{Pradhan} by introducing a complex bosonic
field $\Delta_i(\tau)$ and simpler notation,
\begin{equation}
\exp\!\left[
\frac{U}{2N}
\int d\tau\,
c_{\uparrow}^\dagger
c_{\downarrow}^\dagger
c_{\downarrow}
c_{\uparrow}
\right]
=\int \mathcal D\Delta\,\mathcal D\Delta^*\;
\exp\!\left[
-\int d\tau
\left(
\frac{N|\Delta|^2}{2U}
-
\Delta^* c_{\downarrow} c_{\uparrow}
-
\Delta c_{\uparrow}^\dagger c_{\downarrow}^\dagger
\right)
\right].
\end{equation}
After this transformation, the fermionic action becomes quadratic.
Introducing the Nambu spinor
\begin{equation}
\Psi_{\mathbf{k}\alpha} = 
\begin{pmatrix}
c_{\mathbf{k}\alpha\uparrow} \\
c_{-\mathbf{k}\alpha\downarrow}^\dagger
\end{pmatrix}
\end{equation}
the inverse fermionic Green's function is
\begin{equation}
\mathcal G^{-1}(i\omega_n,\mathbf k)=
\begin{pmatrix}
i\omega_n-\epsilon_{\mathbf k}-i\theta & \Delta \\
\Delta^* & i\omega_n+\epsilon_{\mathbf k}+i\theta
\end{pmatrix}.
\end{equation}
Its determinant yields the quasiparticle spectrum
\begin{equation}
E_{\mathbf k}
=
\sqrt{\epsilon_{\mathbf k}^2+|\Delta|^2},
\end{equation}
which governs the BCS--BEC crossover physics.

Since the fermionic action is quadratic, the Grassmann fields can be
integrated out and the effective action is \cite{Taylor}
\begin{equation}
S_{\mathrm{eff}}[\Delta,\theta]
=
\int_0^\beta d\tau
\frac{N|\Delta|^2}{2U}
-
N\,\mathrm{Tr}\,\ln \mathcal G^{-1}[\Delta,\theta]
\end{equation}
where the trace is taken over lattice sites, imaginary time (or
Matsubara frequencies), Nambu indices, and internal degrees of freedom. The static saddle-point configuration is
$\Delta_i(\tau)=\Delta$.
The trace over Nambu space yields
\begin{equation}
\det \mathcal G^{-1}(i\omega_n,\mathbf k)
=
(i\omega_n-i\theta)^2
-
\left(
\epsilon_{\mathbf k}^2 + |\Delta|^2
\right),
\end{equation}
and the functional trace can therefore be written as
\begin{equation}
\mathrm{Tr}\,\ln \mathcal G^{-1}
=
\sum_{\mathbf k}
\sum_{n}
\ln\!\left[
(i\omega_n-i\theta)^2 - E_{\mathbf k}^2
\right].
\end{equation}
Performing the Matsubara frequency sum using contour-integration gives
\begin{equation}
\sum_{n}
\ln\!\left[(i\omega_n-i\theta)^2 - E_{\mathbf k}^2\right]=\beta E_{\mathbf k}+
\left[\ln\!\left(1+e^{-\beta(E_{\mathbf k}+i\theta)}\right)
+\ln\!\left(1+e^{-\beta(E_{\mathbf k}-i\theta)}\right)
\right]
\end{equation}
Substituting this result back into the effective action yields
\begin{equation}
S_{\mathrm{eff}}[\Delta,\theta]
= N
\Big[
\frac{|\Delta|^2}{2U}
-\sum_{\mathbf k}
\Big(E_{\mathbf k}
+
\left[
\ln\!\left(1+e^{-\beta(E_{\mathbf k}+i\theta)}\right)
+\ln\!\left(1+e^{-\beta(E_{\mathbf k}-i\theta)}\right)
\right]\Big)\Big].
\end{equation}

In the thermodynamic limit, the lattice momentum sum is replaced by an
integral over the first Brillouin zone,
\begin{equation}
\sum_{\mathbf k}
\;\longrightarrow\;
\int_{\mathrm{BZ}}
\frac{d^{d-1} k}{(2\pi)^{d-1}},
\end{equation}
The effective action density then becomes
\begin{equation}
S_{\mathrm{eff}}
=
N
\Big[
\frac{|\Delta|^2}{2U}-\int_{\mathrm{BZ}}
\frac{d^{d-1} k}{(2\pi)^{d-1}}
\Big(E_{\mathbf k}+
\left[
\ln\!\left(1+e^{-\beta(E_{\mathbf k}+i\theta)}\right)
+
\ln\!\left(1+e^{-\beta(E_{\mathbf k}-i\theta)}\right)
\right]\Big)\Big].
\end{equation}

\section{Gap Equations}

The first gap equation of the model for gap $\Delta$ is

\begin{equation}
\frac{\partial S_{\mathrm{eff}}}{\partial \Delta^*} = 0
\end{equation}
and yields the equation
\begin{equation}
\frac{1}{U}
=
\int_{\mathrm{BZ}}
\frac{d^{d-1} k}{(2\pi)^{d-1}}
\,
\frac{1}{E_{\mathbf k}}
\left[
1
-
f(E_{\mathbf k}+i\theta)
-
f(E_{\mathbf k}-i\theta)
\right],
\end{equation}
where 
$f(x) = \frac{1}{e^{\beta x}+1}$.
At zero temperature this reduces to
\begin{equation}
\frac{1}{U}
=
\int_{\mathrm{BZ}}
\frac{d^{d-1} k}{(2\pi)^{d-1}}
\,
\frac{1}{E_{\mathbf k}},
\end{equation}
while for \(\theta=0\) we recover the standard BCS gap equation
\begin{equation}
\frac{1}{U}
=
\int_{\mathrm{BZ}}
\frac{d^{d-1} k}{(2\pi)^{d-1}}
\,
\frac{1-2f(E_{\mathbf k})}{E_{\mathbf k}}.
\end{equation}
The canonical constraint is enforced by the saddle-point condition
with respect to $\theta$:
\begin{equation}
\frac{\partial S_{\mathrm{eff}}}{\partial \theta} = 0.
\end{equation}
This yields the number equation at half-filling
\begin{equation}
N_f
=
\int_{\mathrm{BZ}} \frac{d^{d-1} k}{(2\pi)^{d-1}} \,
\Big[
f(E_{\mathbf k}-i\theta) - f(E_{\mathbf k}+i\theta)
\Big],
\label{N_f}
\end{equation}
The gap equation for $\Delta$ may split into two parts,
\begin{equation}
I_1=\int_{\mathrm{BZ}}
\frac{d^{d-1} k}{(2\pi)^{d-1}}
\,
\frac{1}{E_{\mathbf k}}
\end{equation}
and 
\begin{equation}
I_2=-\int_{\mathrm{BZ}}
\frac{d^{d-1} k}{(2\pi)^{d-1}}
\,
\frac{1}{E_{\mathbf k}}
\left[
f(E_{\mathbf k}+i\theta)
+
f(E_{\mathbf k}-i\theta)
\right]
\end{equation}
For $1D$ chains at finite temperature we take simply
\begin{equation}
I_1=\int_{\mathrm{BZ}}
\frac{dk}{2\pi}\frac{1}{E_{\mathbf k}}
\end{equation}
and
\begin{equation}
I_2=-\int_{\mathrm{BZ}}
\frac{dk}{2\pi}\frac{1}{E_{\mathbf k}}\left[
f(E_{\mathbf k}+i\theta)
+
f(E_{\mathbf k}-i\theta)
\right]
\end{equation}
If we define the vacuum (critical) coupling by
\begin{equation}
\frac{1}{U_c}\equiv I_1(\Delta=0)
=\int_{\mathrm{BZ}}\frac{dk}{2\pi}\frac{1}{|\epsilon_k|}.
\end{equation}
and subtracting this value, since there is a logarithmic divergent part at the edges of Brillouin zone, the renormalized gap equation becomes
\begin{equation}
\frac{1}{U}-\frac{1}{U_c}
=
\int_{\mathrm{BZ}}\frac{dk}{2\pi}
\left(
\frac{1}{\sqrt{\epsilon_k^2+\Delta^2}}-\frac{1}{|\epsilon_k|}
\right)
+ I_2(\Delta,\theta).
\end{equation}
The sign of $\frac{1}{U}-\frac{1}{U_c}$ controls the crossover:
positive corresponds to the BCS regime, negative to the BEC regime, and
$\frac{1}{U}=\frac{1}{U_c}$ marks the crossover (unitarity) point.

\section{Phase Structure: BCS, Unitarity, and BEC Regimes}

\subsection{General Setup}

Using the identity \cite{Roberge}
\begin{equation}
f(E_k+i\theta)+f(E_k-i\theta)=1-\frac{\sinh(\beta E_k)}{\cosh(\beta E_k)+\cos(\beta\theta)},
\end{equation}
and $\phi \equiv \beta\theta$ as a dimensionless variable, the gap equation becomes:
\begin{equation}
\delta_u \equiv \frac{1}{U} - \frac{1}{U_c}
= \int_{\mathrm{BZ}} \frac{dk}{2\pi} \left[ 
\frac{\sinh(\beta E_k)}{E_k\big[\cosh(\beta E_k)+\cos\phi\big]}
- \frac{1}{|\epsilon_k|}
\right]
\end{equation}
For a lattice dispersion $\epsilon_k = -2t\cos k$ (half‑filling at $k=\pm\pi/2$), near the Fermi points $k=\pi/2+q$ and $k=-\pi/2+q$:
\begin{equation}
\epsilon_k \approx \pm 2t q, \qquad 
E_k \approx \sqrt{(2t q)^2 + \Delta^2} \equiv E_q.
\end{equation}
If we split the integral into contributions near $\pm\pi/2$ (within a cutoff $|q|<\Lambda$) and the remainder $C_{\mathrm{UV}}$ (higher momentum part inside BZ, where UV has only symbolic character):
\begin{equation}
\delta_u = \underbrace{\int_{0}^{\Lambda} \frac{dq}{\pi} \left[ 
\frac{\sinh(\beta E_q)}{E_q\big[\cosh(\beta E_q)+\cos\phi\big]}
- \frac{1}{2tq}
\right]}_{\displaystyle \mathcal{F}_{\mathrm{low}}(\Delta,T,\phi;\Lambda)}
\;+\; C_{\mathrm{UV}}(\Lambda)
\end{equation}
and
\begin{equation}
\frac{1}{U_c}\equiv I_1(\Delta=0)
=\int_{0}^{\Lambda}\frac{dq}{\pi}\frac{1}{2tq}.
\end{equation}
Since $C_{\mathrm{UV}}(\Lambda)$ comes from the rest of the Brillouin zone it is \textbf{independent of $\Delta$ and $T$} in the low‑energy limit ($\Lambda \ll 1$, low $T$ and $\Delta$).
The thermal kernel we define is
\begin{equation}
g(x, \phi) \equiv \frac{\sinh x}{\cosh x + \cos\phi}, \qquad x = \beta E_q.
\label{original}
\end{equation}
If we use the key identity
\begin{equation}
g(x,\phi) - 1 = -\dfrac{e^{-x} + \cos\phi}{\cosh x + \cos\phi}
\label{key}
\end{equation}
hence
\begin{equation}
g(x,\phi) \gtreqless 1 \quad \Longleftrightarrow \quad \cos\phi \lesseqgtr -e^{-x}.
\end{equation}
This inequality depends on $x$, and therefore on $q$ through $E_q$. Since $\phi$ is constant, the sign of $g-1$ can change with $q$. At angle $\phi=2\pi/3$ (and similarly at $4\pi/3$), where $\cos\phi=-1/2$, the kernel becomes
\begin{equation}
g(x,2\pi/3) = \frac{\sinh x}{\cosh x - \tfrac{1}{2}},
\end{equation}
and satisfies
\begin{equation}
g(x,2\pi/3)-1 = -\frac{e^{-x}-\tfrac{1}{2}}{\cosh x - \tfrac{1}{2}}.
\end{equation}
This expression changes sign at $x_*=\ln 2$. Consequently, the integrand of the gap equation exhibits a nontrivial structure: energy modes ($x<x_*$) contribute with $g<1$, while-energy modes ($x>x_*$) contribute with $g>1$.

At the critical point $\delta_u=0$, these contributions balance exactly after inclusion of the high energy term $C_{\mathrm{UV}}$. This cancellation is not due to a symmetry of the integrand, but rather reflects a redistribution of spectral weight between low- and high-energy scales controlled by the imaginary chemical potential.

Therefore, the angles $\phi=2\pi/3$ and $4\pi/3$ mark the points where the thermal kernel produces a sign-changing contribution across energy scales. This leads to a particularly sensitive balance in the gap equation at the critical coupling.
Interestingly, as shown in diagrams above, the phase transitions of the model depend on the values of $\delta_u$, $\theta/\pi$ and the sign of thermal kernel. The colors show the sign of $F+C_{\text{UV}}$ but not the coupling regime, since crossover is controlled by $\delta_u$. The blue area says that thermal fluctuations enhance pairing ($g(x,\phi)<1$) at low energies (above $x*$) and the green area says that thermal fluctuations suppress pairing ($g(x,\phi)>1$) still at low energies (but over $x*$). The red dots are exactly on $2\pi/3,4\pi/3$ where $\Delta=0$ and $N_f$ is maximized. The first diagram in figure $1$ was made for $\delta_u=0$ and also for specific values of $\delta_u=+0.3, 0, -0.3$ and for $\delta_u=+0.6, 0, -0.6$ we take the diagrams from figures $2$ and $3$.

\begin{figure}[h]
               \begin{center}
                \includegraphics[scale=0.50]{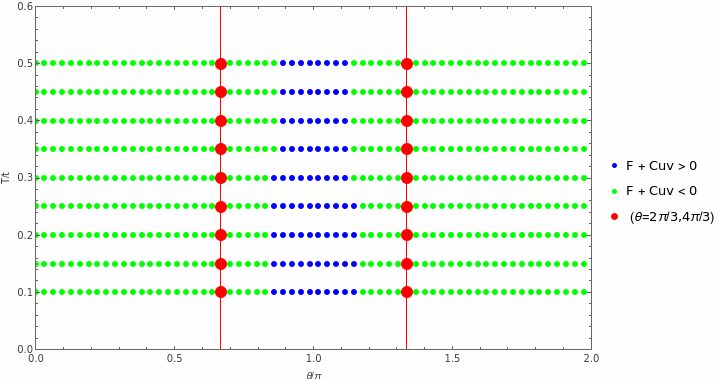}
                \end{center}
                 \caption{BCS-BEC crossover in the $\theta$-$T$ plane showing the thermal window. The blue region indicates enhanced pairing ($g>1$), green indicates suppressed pairing ($g<1$), and red dots at $\beta\theta=2\pi/3,4\pi/3$ mark where $\Delta=0$ and $N_f$ is maximized at unitarity. The parameters we have used in Mathematica plot are $t=1$, $\Lambda=0.5$, $C_{\text{UV}}=-0.2$ for all diagrams.}
                 \label{fig1} 
                 \end{figure}

\begin{figure}[h]
               \begin{center}
                \includegraphics[scale=0.35]{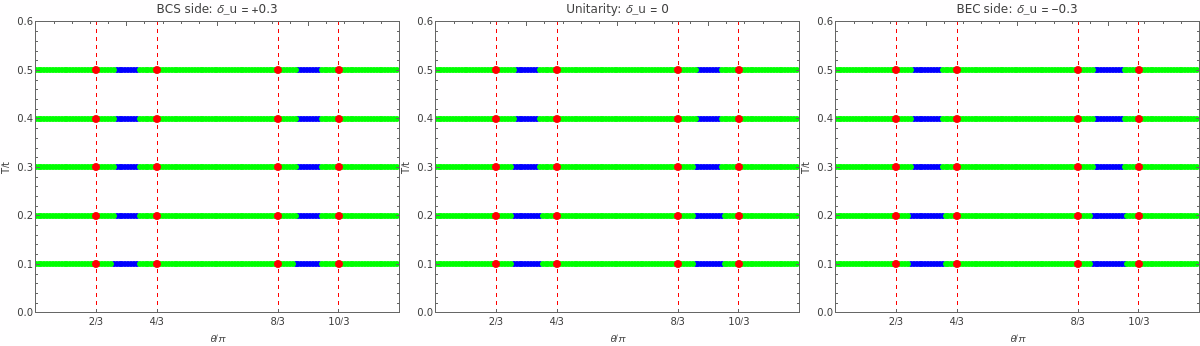}
                \end{center}
                 \caption{BCS-BEC crossover for $\delta_u=+0.3, 0, -0.3$ showing the evolution of the phase boundary. The thermal window at $\beta\theta=2\pi/3,4\pi/3$ remains fixed as $\delta_u$ varies, demonstrating the universal nature of these special angles.}
                 \label{fig2} 
                 \end{figure}
                 
\begin{figure}[h]
               \begin{center}
                \includegraphics[scale=0.35]{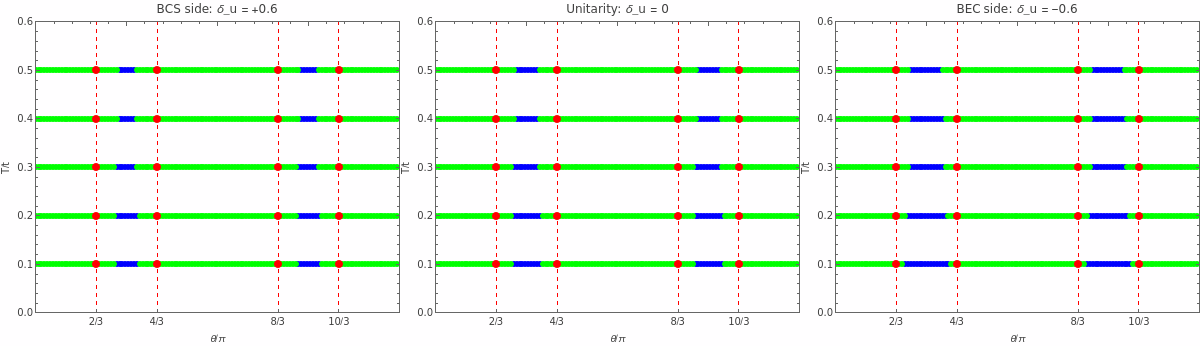}
                \end{center}
                 \caption{BCS-BEC crossover for $\delta_u=+0.6, 0, -0.6$ confirming that the thermal window boundaries are independent of coupling strength. The red dots at $\beta\theta=2\pi/3,4\pi/3$ persist as the points where the gap vanishes at unitarity. We see that the blue dots become more numerous and dominate the phase diagram inside the thermal window at $\delta_u<0$.}
                 \label{fig3} 
                 \end{figure}
Although the diagrams provide a visual representation of the phase thermal window of BCS-BEC crossover, the following analysis in (\ref{sec:Th}) will give an exact calculation of these specific angles.

\subsection{Thermal window at unitarity}                 
\label{sec:Th}

The parameter $\delta_u$ at zero gap is (if we assume for simplicity that $C_{\text{UV}}\approx 0$)
\begin{equation}
 \delta_u^{0}= \frac{1}{U} - \frac{1}{U_c} = \mathcal{F}_{\mathrm{low}}(0,T,\phi),
\end{equation}
where $\mathcal{F}_{\mathrm{low}}$ is evaluated at $\Delta=0$.

Explicitly, setting $E_q \to |2tq|$ in the low-energy integral, we obtain
\begin{equation}
\mathcal{F}_{\mathrm{low}}(0,T,\phi)
=
\int_{0}^{\Lambda} \frac{dq}{\pi}
\left[
\frac{g(\beta |2tq|,\phi)}{2tq}
- \frac{1}{2t q}
\right],
\end{equation}
so that
\begin{equation}
\delta_u^{0} \sim \frac{1}{U} - \frac{1}{U_c}
=\int_{0}^{\Lambda} \frac{dq}{2\pi t\,q}
\big[g(\beta |2tq|,\phi)-1\big].
\label{eq:a2_kernel}
\end{equation}
The sign of $\delta_u^{0}$ is therefore controlled by the kernel $g(x,\phi)-1$, whose structure determines whether low-energy or higher-energy modes dominate the integral.
At the special angle $\phi=2\pi/3$ (and similarly at $4\pi/3$), where $\cos\phi=-1/2$, the kernel satisfies (\ref{key})
which changes sign at the finite value $x_*=\ln 2$. As a result, the integral in Eq.~(\ref{eq:a2_kernel}) receives competing contributions:

\begin{itemize}
\item Energies ($x<x_*$): $g(x,\phi)<1$, giving a \textit{negative} contribution,
\item Energies ($x>x_*$): $g(x,\phi)>1$, giving a \textit{positive} contribution.
\end{itemize}
At the critical coupling $U=U_c$, the parameter $\delta_u^{0}$ is zero so,
\begin{equation}
0=\int_{0}^{\Lambda} \frac{dq}{2\pi t\,q}
\big[g(\beta |2tq|,\phi)-1\big]\rightarrow g(\beta |2tq|,\phi)=1.
\end{equation}
We may say that the sign of $\delta_u^{0}$ depends on how the spectral weight is distributed across energy scales.
In particular, at $\phi=2\pi/3$ and $4\pi/3$, the presence of a sign change in $g(x,\phi)-1$ within the thermal window leads to an easily changeable balance between lower and higher energy contributions. So this makes these angles important.
At $\phi = 2\pi/3,\,4\pi/3$, the competition of low and high energy physics is more clear, since the crossover between $g(x,\phi)<1$ and $g(x,\phi)>1$ occurs at $x_*=\ln 2$. This identifies these angles as points where thermal fluctuations most effectively compete with the interaction scale $1/U - 1/U_c$, controlling the stability of the normal state. If $C_{\text{UV}}$ is negative (without high energy contribution from thermal kernel) then if $\delta_u=0$ we must have $U>U_c$ in order to cancel that contribution and if $C_{\text{UV}}$ is positive (with high energy contribution from thermal kernel) then if $\delta_u=0$ we must have $U<U_c$ in order to cancel that contribution. On the other hand if we consider that $C_{\text{UV}}$ is a shift to $\delta_u$ like
\begin{equation}
\delta_u - C_{\mathrm{UV}} = \mathcal{F}_{\mathrm{low}}(\Delta,T,\phi)
\end{equation}
we take the regimes:
\begin{itemize}
    \item $\delta_u - C_{\mathrm{UV}} > 0$: Requires $\mathcal{F}_{\mathrm{low}} > 0$. This tends to favour $\cos\phi$ small or negative to enhance $g$ at low $q$.
    \item $\delta_u - C_{\mathrm{UV}} < 0$: Requires $\mathcal{F}_{\mathrm{low}} < 0$. This occurs naturally at large $\Delta$ (BEC) or if $\cos\phi \approx 1$ and $T$ small (since $g<1$ suppresses the integrand).
\end{itemize} 
Interestingly, the condition $\cos\phi=-\frac{1}{2}$ plays a role analogous to the appearance of $\ln 2$ in the Gross–Neveu model at imaginary chemical potential \cite{Filothodoros, Filothodoros2}. In both cases, the phase boundary is determined by a balance between interaction energy and a universal thermal contribution arising from low-energy fermionic modes. But here, while in the Gross–Neveu model this contribution appears explicitly as the free energy density, in our formulation it emerges through the structure of the thermal kernel, which maximally enhances low-energy spectral weight at $2\pi/3$.

\subsection{Extrema of $N_f(\phi)$ at $\phi=2\pi/3,\,4\pi/3$}

Introducing the density of states, the fermion number is \footnote{Note that the expression in (\ref{N_f}) is purely imaginary. The physical fermion number is obtained by multiplying by $-i$, yielding a real quantity that measures the imbalance between particle-like and hole-like excitations.}
\begin{equation}
N_f(\phi,T)
=
\int_{-2t}^{2t} d\epsilon \, \rho(\epsilon)\, \frac{\sin\phi}{\cosh(\beta|\epsilon|) + \cos\phi},
\label{eq:Nf_correct}
\end{equation}
where we have used the identity:
\begin{equation}
f(E - i\phi) - f(E + i\phi) = \frac{i \sin\phi}{\cosh(\beta E) + \cos\phi},
\end{equation}
We assume particle--hole symmetry of the band,
\begin{equation}
\rho(\epsilon)=\rho(-\epsilon).
\end{equation}
Note that the integrand depends on $|\epsilon|$, so it is an \textbf{even} function of $\epsilon$.
Differentiating \(N_f\) with respect to $\phi$:
\begin{equation}
\frac{dN_f}{d\phi}=
\cos\phi \int_{-2t}^{2t} d\epsilon\,\rho(\epsilon) \frac{1}{\cosh(\beta|\epsilon|) + \cos\phi}
\;+\;\sin\phi \frac{\partial}{\partial\phi} \int_{-2t}^{2t} d\epsilon\,\rho(\epsilon) \frac{1}{\cosh(\beta|\epsilon|) + \cos\phi}.
\end{equation}
If we define:
\begin{equation}
I(\phi) \equiv \int_{-2t}^{2t} d\epsilon\,\rho(\epsilon) \frac{1}{\cosh(\beta|\epsilon|) + \cos\phi}.
\end{equation}
Then:
\begin{equation}
\frac{dN_f}{d\phi} = \cos\phi \, I(\theta) + \sin\phi \, I'(\phi).
\end{equation}
Now if we compute \(I'(\phi)\):
\begin{equation}
I'(\phi) =\sin\phi \int_{-2t}^{2t} d\epsilon\,\rho(\epsilon) \frac{1}{(\cosh(\beta|\epsilon|) + \cos\phi)^2}.
\end{equation}
and define
\begin{equation}
J(\phi) \equiv \int_{-2t}^{2t} d\epsilon\,\rho(\epsilon) \frac{1}{(\cosh(\beta|\epsilon|) + \cos\phi)^2}.
\end{equation}
Thus:
\begin{equation}
\frac{dN_f}{d\phi} = \cos\phi \, I(\phi) + \sin^2\phi \, J(\phi).
\label{eq:dNf}
\end{equation}
Both $I(\phi)$ and $J(\phi)$ are \textbf{even} functions of $\phi$ (they depend only on $\cos\phi$), and the integrands are even in $\epsilon$ (since they depend on $|\epsilon|$), so the integrals are nonzero in general.

Setting $\frac{dN_f}{d\phi} = 0$ gives:
\begin{equation}
\cos\phi \, I(\phi) + \sin^2\phi \, J(\phi) = 0.
\label{eq:extremum}
\end{equation}
This equation determines the stationary points of $N_f(\phi)$.
At $\phi = 2\pi/3$ and $4\pi/3$, we have:
\begin{equation}
\cos\phi = -\frac{1}{2}, \qquad \sin^2\phi = \frac{3}{4}.
\end{equation}
Substituting into Eq.~(\ref{eq:extremum}):
\begin{equation}
-\frac{1}{2} \, I\!\left(\frac{2\pi}{3}\right) + \frac{3}{4} \, J\!\left(\frac{2\pi}{3}\right) = 0.
\end{equation}
This implies:
\begin{equation}
J\!\left(\frac{2\pi}{3}\right) = \frac{2}{3} \, I\!\left(\frac{2\pi}{3}\right).
\end{equation}
Differentiating Eq.~(\ref{eq:dNf}) once more:
\begin{align}
\frac{d^2N_f}{d\phi^2}&= -\sin\phi \, I(\phi) + \cos\phi \, I'(\phi) +2\sin\phi\cos\phi \, J(\phi) + \sin^2\phi \, J'(\phi) \nonumber \\
&= -\sin\phi \, I(\phi) + 3\sin\phi\cos\phi \, J(\phi) + \sin^2\phi \, J'(\phi).
\end{align}
At $\phi = 2\pi/3$, $\sin\phi = \frac{\sqrt{3}}{2} \neq 0$, and using the relation $J = \frac{2}{3} I$:
\begin{equation}
\frac{d^2N_f}{d\phi^2}\bigg|_{\phi=2\pi/3}= -\sqrt{3} \, I + \frac{3}{4} J'.
\end{equation}
The function $I(\phi)$ is strictly positive, since its integrand is positive definite. The sign of $J'(\phi)$ depends on the details of the spectrum. Therefore we could say that the sign of second derivative depends on the competition between the two terms.
The angles $\phi=2\pi/3$ and $4\pi/3$ do not correspond to universal inflection points of $N_f(\phi)$. Instead, they are special because the kernel
\begin{equation}
\frac{1}{\cosh(\beta|\epsilon|)+\cos\phi}
\end{equation}
is critical at low energies when $\cos\phi=-1/2$, which as we know it controls the sign of the quadratic coefficient in the Landau expansion (\ref{Appendix C}). The extrema and curvature of $N_f(\phi)$ near these angles are determined dynamically and must be evaluated numerically or within controlled approximations. Using a standard approximation in low-temperature expansions (\ref{Appendix B}) we take

\begin{itemize}
\item At $\phi = 2\pi/3$: $\displaystyle \frac{dN_f}{d\phi}=0$ and
$\displaystyle \frac{d^2N_f}{d\phi^2}<0$, therefore $N_f$ has a
\emph{local maximum}. This corresponds to the maximal particle-hole
asymmetry with particle-like excitations dominating.
\item At $\phi = 4\pi/3$: $\displaystyle \frac{dN_f}{d\phi}=0$ and
$\displaystyle \frac{d^2N_f}{d\phi^2}>0$, therefore $N_f$ has a
\emph{local minimum}. This corresponds to the maximal (negative)
particle-hole asymmetry with hole-like excitations dominating.
\end{itemize}

These results are consistent with the physical expectation that
$N_f(\phi)$ is an odd function of $\phi$ about $\phi=\pi$,
$N_f(\pi+\delta) = -N_f(\pi-\delta)$, so a maximum at $2\pi/3$ implies
a minimum at $4\pi/3$.

\paragraph{Phase structure and the special angle $2\pi/3$.}

It is extremely interesting that the condition $\cos\phi = -\frac{1}{2}$ arises in our formulation through the thermal kernel (\ref{original}).
We obtain that at specific angles, corresponding to $\phi = 2\pi/3$ and $4\pi/3$, the kernel leads to a nontrivial redistribution of spectral weight across energy scales. In particular, it seems like the sign of $g(x,\phi) - 1$ depends on the competition between $e^{-x}$ and $|\cos\phi| = 1/2$. This implies that low-energy and high-energy contributions enter the gap equation with opposite signs. As a result, the integral determining the interaction shift $\delta_u = 1/U - 1/U_c$ is governed by a balance between these regions. Physically talking, this identifies $\phi = 2\pi/3,\,4\pi/3$ as crossover angles where the thermal kernel changes character: low-energy modes are enhanced while higher-energy contributions are suppressed and this affects the sign of $\delta_u$.

Notably, $\phi=2\pi/3$ corresponds to the primitive third root of unity $e^{i2\pi/3}$, suggesting that the three crossover regimes
(BCS, unitarity, and BEC) may be naturally organized by interpolation between three phase regimes of the fermionic boundary condition.

Interestingly, the same phase value appears in several other systems with $U(1)$ phases. For example, lattice models with magnetic flux $\phi=2\pi/3$ per plaquette exhibit nontrivial Berry curvature
and topological band structures \cite{Liu}. Also, cold-atom quantum walks \cite{Li} and Floquet-engineered gauge fields \cite{Wang} can realize tunneling phases close
to $2\pi/3$, modifying the interference of hopping paths and the resulting quasienergy spectra. In those systems the phase acts in
real space or momentum space, but in our case it appears as a twist along the imaginary-time direction.

Although the microscopic physics of these systems are different, the angle $2\pi/3$ is special. We may say that it reflects an algebraic structure with interpolation between three phase regimes.
In our thermal problem this shows up as a reorganization of the
kernel governing particle–hole excitations. This phenomenon produces the behavior of the fermion number $N_f$ and the gap equation near the unitarity point.

\subsection{Exact values of $N_f$ as functions of ratio $\frac{T}{t}$ for small $T$}

\label{sec:Nf_value}

We now compute explicitly the fermion number $N_f$ at the special angles $\phi = 2\pi/3$ and $4\pi/3$. Starting from Eq.~(\ref{eq:Nf_correct}):
\begin{equation}
N_f(\phi,T) = \int_{-2t}^{2t} d\epsilon \, \rho(\epsilon) \, \frac{\sin\phi}{\cosh(\beta|\epsilon|) + \cos\phi},
\label{eq:Nf_start}
\end{equation}
with the 1D density of states $\rho(\epsilon) = \frac{1}{\pi\sqrt{4t^2 - \epsilon^2}}$ for $|\epsilon| \le 2t$. At $\phi = 2\pi/3$, we have $\sin(2\pi/3) = \frac{\sqrt{3}}{2}$ and $\cos(2\pi/3) = -\frac{1}{2}$. Using the fact that the integrand is even we take:

\begin{equation}
N_f\!\left(\frac{2\pi}{3},T\right) = \sqrt{3} \int_{0}^{2t} d\epsilon \, \rho(\epsilon) \, \frac{1}{\cosh(\beta\epsilon) - \frac{1}{2}}.
\label{eq:Nf_2pi3_start}
\end{equation}
Introduce the variable $x = \beta\epsilon$. Then $\epsilon = x/\beta$, $d\epsilon = dx/\beta$, and:

\begin{equation}
\rho(\epsilon) = \frac{1}{\pi\sqrt{4t^2 - (x/\beta)^2}} = \frac{1}{2\pi t \sqrt{1 - (x/(2t\beta))^2}}.
\label{eq:rho_x}
\end{equation}
Thus:

\begin{equation}
N_f\!\left(\frac{2\pi}{3},T\right) = \frac{\sqrt{3}}{2\pi t\beta} \int_{0}^{2t\beta} \frac{dx}{\sqrt{1 - (x/(2t\beta))^2}} \, \frac{1}{\cosh x - \frac{1}{2}}.
\label{eq:Nf_2pi3_int}
\end{equation}
If we use
\begin{equation}
I = \int_{0}^{\infty} \frac{dx}{\cosh x - \frac{1}{2}} = \frac{4\pi}{3\sqrt{3}},
\label{eq:exact_integral}
\end{equation}
which follows from the identity $\int_{0}^{\infty} \frac{dx}{\cosh x + a} = \frac{2}{\sqrt{1-a^2}} \arctan\left( \sqrt{\frac{1-a}{1+a}} \right)$ for $|a| < 1$, specialized to $a = -1/2$ and insert this result in Eq.~(\ref{eq:Nf_2pi3_int}) with the approximation $\sqrt{1 - (x/(2t\beta))^2} \approx 1$ for the dominant small-$x$ region:

\begin{equation}
N_f\!\left(\frac{2\pi}{3},T\right) \approx \frac{\sqrt{3}}{2\pi t\beta} \int_{0}^{\infty} \frac{dx}{\cosh x - \frac{1}{2}} = \frac{\sqrt{3}}{2\pi t\beta} \cdot \frac{4\pi}{3\sqrt{3}} = \frac{2}{3t\beta}.
\label{eq:Nf_2pi3_final}
\end{equation}
Since $\beta = 1/T$, we obtain the simple linear temperature dependence:

\begin{equation}
\boxed{ N_f\!\left(\frac{2\pi}{3},T\right) = \frac{2}{3} \cdot \frac{T}{t} }.
\label{eq:Nf_2pi3_result}
\end{equation}
At $\theta = 4\pi/3$, we have $\sin(4\pi/3) = -\frac{\sqrt{3}}{2}$ and $\cos(4\pi/3) = -\frac{1}{2}$. The calculation proceeds identically, yielding:

\begin{equation}
N_f\!\left(\frac{4\pi}{3},T\right) = -\frac{\sqrt{3}}{2\pi t\beta} \int_{0}^{\infty} \frac{dx}{\cosh x - \frac{1}{2}} = -\frac{2}{3t\beta} = -\frac{2}{3} \cdot \frac{T}{t}.
\label{eq:Nf_4pi3_result}
\end{equation}
Thus the fermion number at the special angles is simply:

\begin{equation}
N_f(\phi,T) = \frac{2}{3} \cdot \frac{T}{t} \times \begin{cases}
+1 & \phi = 2\pi/3 \\
-1 & \phi = 4\pi/3
\end{cases}
\label{eq:Nf_both}
\end{equation}

\subsection{Exact values of $N_f$ as functions of ratio $\frac{T}{t}$ for large $T$}
\label{sec:Nf_highT}

We now examine the behavior of $N_f$ at angle $\phi = 2\pi/3$ in the opposite limit of high temperature, i.e., small $\beta = 1/T$. Starting from Eq.~(\ref{eq:Nf_correct}):

\begin{equation}
N_f(\phi,T) = \int_{-2t}^{2t} d\epsilon \, \rho(\epsilon) \, \frac{\sin\phi}{\cosh(\beta|\epsilon|) + \cos\phi},
\label{eq:Nf_highT_start}
\end{equation}
At $\phi = 2\pi/3$, $\sin(2\pi/3) = \frac{\sqrt{3}}{2}$ and $\cos(2\pi/3) = -\frac{1}{2}$, giving:

\begin{equation}
N_f\!\left(\frac{2\pi}{3},T\right) = \frac{\sqrt{3}}{2} \int_{-2t}^{2t} d\epsilon \, \rho(\epsilon) \, \frac{1}{\cosh(\beta|\epsilon|) - \frac{1}{2}}.
\label{eq:Nf_2pi3_highT}
\end{equation}
In the high-temperature limit $\beta \to 0$, we have $\beta|\epsilon| \ll 1$ we expand $\cosh(\beta|\epsilon|)$:

\begin{equation}
\cosh(\beta|\epsilon|) = 1 + \frac{(\beta|\epsilon|)^2}{2} + \frac{(\beta|\epsilon|)^4}{24} + \mathcal{O}(\beta^6).
\label{eq:cosh_expand}
\end{equation}
Thus the denominator becomes:

\begin{equation}
\cosh(\beta|\epsilon|) - \frac{1}{2} = \frac{1}{2} + \frac{(\beta|\epsilon|)^2}{2} + \frac{(\beta|\epsilon|)^4}{24} + \cdots.
\label{eq:denom_expand}
\end{equation}
To leading order in small $\beta$:

\begin{equation}
\frac{1}{\cosh(\beta|\epsilon|) - \frac{1}{2}} \approx \frac{1}{\frac{1}{2} + \frac{(\beta|\epsilon|)^2}{2}} = \frac{2}{1 + (\beta|\epsilon|)^2}.
\label{eq:frac_approx}
\end{equation}
Substituting this approximation into Eq.~(\ref{eq:Nf_2pi3_highT}):

\begin{equation}
N_f\!\left(\frac{2\pi}{3},T\right) \approx \frac{\sqrt{3}}{2} \int_{-2t}^{2t} d\epsilon \, \rho(\epsilon) \, \frac{2}{1 + (\beta|\epsilon|)^2} =\sqrt{3} \int_{-2t}^{2t} d\epsilon \, \rho(\epsilon) \, \frac{1}{1 + (\beta|\epsilon|)^2}.
\label{eq:Nf_approx1}
\end{equation}
Introduce the dimensionless variable $u = \beta\epsilon$. Then $\epsilon = u/\beta$, $d\epsilon = du/\beta$, and the density of states becomes:

\begin{equation}
\rho(\epsilon) = \frac{1}{\pi\sqrt{4t^2 - \epsilon^2}} = \frac{1}{\pi\sqrt{4t^2 - (u/\beta)^2}} = \frac{1}{\pi\sqrt{4t^2 - u^2/\beta^2}}.
\label{eq:rho_u}
\end{equation}
The integration limits become $u = \pm 2t\beta$. Since $\beta$ is small, these limits are small, so the integration range in $u$ is narrow. The integral becomes:

\begin{equation}
N_f\!\left(\frac{2\pi}{3},T\right) \approx \sqrt{3} \int_{-2t\beta}^{2t\beta} \frac{du}{\beta} \, \frac{1}{\pi\sqrt{4t^2 - u^2/\beta^2}} \, \frac{1}{1 + u^2}.
\label{eq:Nf_approx2}
\end{equation}
Factor $2t$ out of the square root:

\begin{equation}
\sqrt{4t^2 - \frac{u^2}{\beta^2}} = 2t \sqrt{1 - \left(\frac{u}{2t\beta}\right)^2}.
\label{eq:sqrt_factor}
\end{equation}
Thus:

\begin{equation}
N_f\!\left(\frac{2\pi}{3},T\right) \approx \frac{\sqrt{3}}{\pi\beta} \int_{-2t\beta}^{2t\beta} \frac{du}{2t\sqrt{1 - (u/(2t\beta))^2}} \, \frac{1}{1 + u^2}.
\label{eq:Nf_approx3}
\end{equation}
Let $v = \frac{u}{2t\beta}$. Then $u = 2t\beta v$, $du = 2t\beta dv$, and the integration limits become $v = \pm 1$. Substituting:

\begin{equation}
N_f\!\left(\frac{2\pi}{3},T\right) \approx \frac{\sqrt{3}}{\pi\beta} \int_{-1}^{1} \frac{2t\beta dv}{2t\sqrt{1-v^2}} \, \frac{1}{1 + (2t\beta v)^2}= \frac{\sqrt{3}}{\pi} \int_{-1}^{1} \frac{dv}{\sqrt{1-v^2}} \, \frac{1}{1 + (2t\beta v)^2}.
\label{eq:Nf_approx4}
\end{equation}
The factors $\beta$ and $t$ cancel nicely, leaving an integral that depends on $\beta$ only through the denominator.
For small $\beta$, we expand the denominator in powers of $\beta^2$:

\begin{equation}
\frac{1}{1 + (2t\beta v)^2} = 1 - (2t\beta v)^2 + (2t\beta v)^4 - (2t\beta v)^6 + \cdots.
\label{eq:series_expand}
\end{equation}
Thus:

\begin{align}
N_f\!\left(\frac{2\pi}{3},T\right)&\approx \frac{\sqrt{3}}{\pi} \int_{-1}^{1} \frac{dv}{\sqrt{1-v^2}} \Big[ 1 - 4t^2\beta^2 v^2 + 16t^4\beta^4 v^4 -64t^6\beta^6 v^6 +\cdots \Big] \nonumber \\
&= \frac{\sqrt{3}}{\pi} \left[ I_0 - 4t^2\beta^2 I_2 + 16t^4\beta^4 I_4 - 64t^6\beta^6 I_6 + \cdots \right],
\label{eq:Nf_series}
\end{align}
where we define the integrals:

\begin{equation}
I_n = \int_{-1}^{1} \frac{v^n dv}{\sqrt{1-v^2}}.
\label{eq:moments_def}
\end{equation}
These integrals are standard and can be evaluated using the substitution $v = \sin\phi$ (\ref{eq:moments_values}).
Substituting these values into Eq.~(\ref{eq:Nf_series}):

\begin{align}
N_f\!\left(\frac{2\pi}{3},T\right) &\approx \frac{\sqrt{3}}{\pi} \Big[ \pi - 4t^2\beta^2 \cdot \frac{\pi}{2} + 16t^4\beta^4 \cdot \frac{3\pi}{8} -64t^6\beta^6 \cdot \frac{5\pi}{16} + \cdots \Big] \nonumber \\
&= \frac{\sqrt{3}}{\pi} \cdot \pi \left[ 1 - 2t^2\beta^2 + 6t^4\beta^4 - 20t^6\beta^6 + \cdots \right] \nonumber \\
&= \sqrt{3} \left[ 1 - 2\left(\frac{t}{T}\right)^2 + 6\left(\frac{t}{T}\right)^4 - 20\left(\frac{t}{T}\right)^6 + \cdots \right].
\label{eq:Nf_highT_final}
\end{align}
where $T = 1/\beta$.
Obviously since $\sin(4\pi/3)=-\frac{\sqrt{3}}{2}$
\begin{equation}
N_f\!\left(\frac{4\pi}{3},T\right)=-N_f\!\left(\frac{2\pi}{3},T\right)
\end{equation} 
The high-temperature expansion reveals several remarkable features:

\begin{itemize}
\item \textbf{Leading term}: As $T \to \infty$, $N_f$ approaches the constant $\sqrt{3} \approx 1.732$, independent of both $t$ and $T$.

\item \textbf{Correction terms}: The expansion is in even powers of $t/T$, with coefficients $2, 6, 20, \ldots$ that follow the pattern from (\ref{Appendix A}). These coefficients are the same as those appearing in the expansion of $\frac{1}{\sqrt{1-x^2}}$, pointing out the geometry of the density of states.

\item \textbf{Crossover scale}: The expansion is valid for $T \gg t$, i.e., when thermal energy dominates over the hopping amplitude. In this regime, thermal fluctuations cancel the details of the lattice, and $N_f$ goes to its high-temperature limit.

\item \textbf{Contrast with low-$T$ limit}: At low temperatures ($T \ll t$), we found $N_f \approx \frac{2}{3} \cdot \frac{T}{t}$, which is linear and small. The crossover of $N_f$ happens when the low-T and high-T expressions are equal. If we solve numerically 
\begin{equation}
\frac{2x}{3}=\sqrt{3}\Big[1-\frac{2}{x^2}\Big]
\end{equation}
with $x=\frac{T}{t}$, we take
\begin{equation}
x=\frac{T}{t}\approx 2
\end{equation}
\end{itemize}
The high-temperature limit $N_f \to \sqrt{3}$ shows that the imaginary chemical potential $\phi = 2\pi/3$ induces a fixed particle-hole asymmetry, independent of material parameters, even at infinite temperature. For example at $T>2t$, fermions beginning from the Fermi-Level, have energies to overcame the Brillouin zone boarders and "hit" the walls of the zone. So now the fermion number cannot keep growing linearly and goes to a specific limit.

\section{Conclusions}

In this work, we have investigated the BCS-BEC crossover of the large $N$ attractive Fermi-Hubbard model on a one-dimensional lattice in the presence of an imaginary chemical potential $i\beta\theta$. The main messages of our work are that the crossover is governed by three key parameters $g(\beta E_k, \beta\theta)$, $\delta_u$ and the imaginary chemical potential $i\theta$ and at unitarity there is a thermal window where superconducting gap $\Delta$ vanishes and the fermion number $N_f$ is maximum/minimum at specific points. Inside this thermal window BCS and BEC regimes dominate the phase. 

Through an approximation at half-filling and near the Fermi points, we find the exact edges of the thermal window at $2\pi/3$ and $4\pi/3$. Finally, we have proven that these angles mark a change in $N_f$ curvature through the condition $\cos\phi=-1/2$. If first derivative vanishes at the edges, then second derivative is negative at $2\pi/3$ and positive at $4\pi/3$ and we may say that they are special points that separate particle-like ($\phi<2\pi/3$) from hole-like ($\phi>4\pi/3$) excitation regimes.

We think that our results offer a new window into the BCS-BEC crossover physics and can be extended in several directions. For example one might be tempted to extend our theory to higher dimensions (for example in $2d$ \cite{Dupuis}) to determine whether the special angles are independent from dimension or not. Possible $2d$ lattice geometries could be the square and honeycomb. Another interesting future research could be the thermal mapping between our model and the repulsive Bose Hubbard model at imaginary chemical potential. Furthermore, it would be interesting to examine the possibility of connecting the imaginary chemical potential with experiment results of Floquet engineering or synthetic gauge fields offers an opportunity to test our results in ultracold atoms.

%
%

\ack{I would like to thank Anastasios Petkou for helpful discussion.}

\data{All data that support the findings of this study are included within the article.}

\section*{Appendix A. Useful integrals}
\label{Appendix A}

Some special integrals are

\begin{align}
I_0 &= \int_{-1}^{1} \frac{dv}{\sqrt{1-v^2}} = \pi, \\
I_2 &= \int_{-1}^{1} \frac{v^2 dv}{\sqrt{1-v^2}} = \frac{\pi}{2}, \\
I_4 &= \int_{-1}^{1} \frac{v^4 dv}{\sqrt{1-v^2}} = \frac{3\pi}{8}, \\
I_6 &= \int_{-1}^{1} \frac{v^6 dv}{\sqrt{1-v^2}} = \frac{5\pi}{16}, \\
I_{2n} &= \pi \cdot \frac{(2n)!}{2^{2n}(n!)^2}, \quad I_{2n+1} = 0.
\label{eq:moments_values}
\end{align}

\section*{Appendix B. Local maximum/minimum of fermion number $N_f$}
\label{Appendix B}
We begin with
\begin{equation}
\frac{d^2N_f}{d\phi^2}
= -\sin\phi \, I(\phi) + 3\sin\phi\cos\phi \, J(\phi) + \sin^2\phi \, J'(\phi).
\label{eq:d2Nf_general}
\end{equation}
\begin{itemize}
\item
At $\phi = 2\pi/3$ we have $\sin\phi = \sqrt{3}/2$, $\cos\phi = -1/2$,
and the extremum condition $\cos\phi \, I + \sin^2\phi \, J = 0$ yields
$J = \frac{2}{3} I$. Substituting these into Eq.~(\ref{eq:d2Nf_general})
gives
\begin{align}
\frac{d^2N_f}{d\phi^2}\bigg|_{\phi=2\pi/3}
&= -\frac{\sqrt{3}}{2} I
+ 3\cdot\frac{\sqrt{3}}{2}\!\left(-\frac{1}{2}\right)\!\left(\frac{2}{3}I\right)
+ \frac{3}{4} J' \notag\\
&= -\frac{\sqrt{3}}{2} I - \frac{\sqrt{3}}{2} I + \frac{3}{4} J'
= -\sqrt{3}\, I + \frac{3}{4} J'.
\label{eq:d2Nf_at_2pi3}
\end{align}
We now compute $J'(\phi)$. Differentiating $J(\phi)$,
\begin{equation}
J'(\phi) = 4\sin\phi \int_{0}^{2t} d\epsilon \, \rho(\epsilon) \,
\frac{1}{\bigl(\cosh(\beta\epsilon) + \cos\phi\bigr)^3}.
\label{eq:Jprime_general}
\end{equation}
At $\phi = 2\pi/3$, $\sin(2\pi/3) = \sqrt{3}/2 > 0$, so
\begin{equation}
J'\!\left(\frac{2\pi}{3}\right)
= 2\sqrt{3} \int_{0}^{2t} d\epsilon \, \rho(\epsilon) \,
\frac{1}{\bigl(\cosh(\beta\epsilon) - \frac{1}{2}\bigr)^3}
> 0.
\label{eq:Jprime_2pi3}
\end{equation}
The integrand is strictly positive because $\rho(\epsilon)>0$ and
$\cosh(\beta\epsilon) - 1/2 \ge 1/2$ for all $\epsilon$. Hence
$J'(2\pi/3)$ is positive.

To determine the sign of $d^2N_f/d\phi^2$ we evaluate the two terms in
Eq.~(\ref{eq:d2Nf_at_2pi3}) explicitly. At low temperatures
($\beta\to\infty$) the integrals are dominated by $\epsilon\approx 0$.
Using the approximations $\rho(\epsilon)\approx 1/(2\pi t)$,
$\cosh(\beta\epsilon)\approx 1+(\beta\epsilon)^2/2$, and extending the
integration to infinity, we obtain
\begin{align}
I &\approx \frac{2}{2\pi t} \int_0^\infty \frac{2}{1+(\beta\epsilon)^2}\,d\epsilon
= \frac{2}{t\beta},\\
J' &\approx \frac{2\sqrt{3}}{2\pi t} \int_0^\infty \frac{8}{\bigl(1+(\beta\epsilon)^2\bigr)^3}\,d\epsilon
= \frac{3\sqrt{3}}{2t\beta}.
\end{align}
Inserting these into Eq.~(\ref{eq:d2Nf_at_2pi3}) gives
\begin{equation}
\frac{d^2N_f}{d\phi^2}\bigg|_{\phi=2\pi/3}
\approx -\sqrt{3}\cdot\frac{2}{t\beta}
+ \frac{3}{4}\cdot\frac{3\sqrt{3}}{2t\beta}
= -\frac{7\sqrt{3}}{8t\beta} < 0.
\end{equation}
Thus, at low temperatures, $\phi = 2\pi/3$ is a \emph{local maximum}
of $N_f(\phi)$.
\item
At $\phi = 4\pi/3$ we have $\sin\phi = -\sqrt{3}/2$, $\cos\phi = -1/2$,
and the same extremum condition gives $J = \frac{2}{3} I$ again.
Substituting into Eq.~(\ref{eq:d2Nf_general}) yields
\begin{align}
\frac{d^2N_f}{d\phi^2}\bigg|_{\phi=4\pi/3}
&= -\left(-\frac{\sqrt{3}}{2}\right) I
+ 3\left(-\frac{\sqrt{3}}{2}\right)\!\left(-\frac{1}{2}\right)\!\left(\frac{2}{3}I\right)
+ \frac{3}{4} J' \notag\\
&= \frac{\sqrt{3}}{2} I + \frac{\sqrt{3}}{2} I + \frac{3}{4} J'
= \sqrt{3}\, I + \frac{3}{4} J'.
\label{eq:d2Nf_at_4pi3}
\end{align}
From Eq.~(\ref{eq:Jprime_general}), at $\phi = 4\pi/3$,
$\sin(4\pi/3) = -\sqrt{3}/2$, so
\begin{equation}
J'\!\left(\frac{4\pi}{3}\right)
= -2\sqrt{3} \int_{0}^{2t} d\epsilon \, \rho(\epsilon) \,
\frac{1}{\bigl(\cosh(\beta\epsilon) - \frac{1}{2}\bigr)^3}
< 0.
\end{equation}
Using the same low-temperature approximations as above,
$I \approx 2/(t\beta)$ and $J' \approx -3\sqrt{3}/(2t\beta)$, we find
\begin{equation}
\frac{d^2N_f}{d\phi^2}\bigg|_{\phi=4\pi/3}
\approx \sqrt{3}\cdot\frac{2}{t\beta}
+ \frac{3}{4}\!\left(-\frac{3\sqrt{3}}{2t\beta}\right)
= \frac{7\sqrt{3}}{8t\beta} > 0.
\end{equation}
Hence $\phi = 4\pi/3$ is a \emph{local minimum} of $N_f(\phi)$.
\end{itemize}

\section*{Appendix C. The $a_2=0$ critical point}
\label{Appendix C}
If one wants to find the critical thermal window at unitarity then using the Landau theory of phase transitions by expanding the thermal potential near a critical point, then:
\begin{equation}
\Omega(\Delta)=\Omega(0)+a_2\Delta^2+a_4\Delta^4+a_6\Delta^6+..
\end{equation}
where $a_2=0$ gives the critical point where the curvature of thermal potential changes sign.
We start from the thermodynamic potential
\begin{equation}
\Omega(\Delta) = \frac{\Delta^2}{2U}
- \int \frac{dk}{2\pi} E_k
- \frac{1}{\beta} \int \frac{dk}{2\pi} \ln\!\left[1 + 2e^{-\beta E_k}\cos\phi + e^{-2\beta E_k}\right],
\label{eq:Omega_start}
\end{equation}
where \(E_k = \sqrt{\epsilon_k^2 + \Delta^2}\). The first derivative with respect to \(\Delta\) is
\begin{equation}
\frac{\partial \Omega}{\partial \Delta}
= \frac{\Delta}{U}
- \int \frac{dk}{2\pi} \frac{\Delta}{E_k}
- \frac{1}{\beta} \int \frac{dk}{2\pi} \frac{1}{X} \frac{\partial X}{\partial \Delta},
\end{equation}
with \(X = 1 + 2e^{-\beta E_k}\cos\phi + e^{-2\beta E_k}\).

We have
\begin{equation}
\frac{\partial X}{\partial \Delta}
= -2\beta \frac{\Delta}{E_k} e^{-\beta E_k}\cos\phi
- 2\beta \frac{\Delta}{E_k} e^{-2\beta E_k}
= -2\beta \frac{\Delta}{E_k} e^{-\beta E_k}\bigl(\cos\phi + e^{-\beta E_k}\bigr).
\end{equation}
Substituting,
\begin{align}
\frac{\partial \Omega}{\partial \Delta}
&=\frac{\Delta}{U} - \int \frac{dk}{2\pi} \frac{\Delta}{E_k} - \frac{1}{\beta} \int \frac{dk}{2\pi} \frac{1}{X}
\left[ -2\beta \frac{\Delta}{E_k} e^{-\beta E_k}\bigl(\cos\phi + e^{-\beta E_k}\bigr) \right] \nonumber\\
&= \frac{\Delta}{U} - \int \frac{dk}{2\pi} \frac{\Delta}{E_k}
+ 2 \int \frac{dk}{2\pi} \frac{\Delta}{E_k}
\frac{e^{-\beta E_k}\bigl(\cos\phi + e^{-\beta E_k}\bigr)}{X}.
\end{align}
Using the identity
\begin{equation}
X = 1 + 2e^{-\beta E_k}\cos\phi + e^{-2\beta E_k}
= 2e^{-\beta E_k}\bigl(\cosh(\beta E_k) + \cos\theta\bigr),
\end{equation}
we obtain
\begin{equation}
\frac{e^{-\beta E_k}\bigl(\cos\phi + e^{-\beta E_k}\bigr)}{X}
= \frac{\cos\phi + e^{-\beta E_k}}{2\bigl(\cosh(\beta E_k) + \cos\phi\bigr)}.
\end{equation}
Thus,
\begin{equation}
\frac{\partial \Omega}{\partial \Delta}
= \frac{\Delta}{U} - \int \frac{dk}{2\pi} \frac{\Delta}{E_k}
+ \int \frac{dk}{2\pi} \frac{\Delta}{E_k}
\frac{\cos\phi + e^{-\beta E_k}}{\cosh(\beta E_k) + \cos\phi}.
\label{eq:first_derivative}
\end{equation}
The second derivative is
\begin{equation}
\frac{\partial^2 \Omega}{\partial \Delta^2}
= \frac{1}{U} - \int \frac{dk}{2\pi} \frac{1}{E_k}
+ \int \frac{dk}{2\pi} \frac{1}{E_k}
\frac{\cos\phi + e^{-\beta E_k}}{\cosh(\beta E_k) + \cos\phi}
+ \Delta \frac{\partial}{\partial \Delta} \bigl[ \cdots \bigr],
\end{equation}
where the term proportional to $\Delta$ is very small for $\Delta\rightarrow 0$.
Hence, the Landau coefficient $a_2(T,\phi) = \frac{\partial^2 \Omega}{\partial \Delta^2}$ is
\begin{equation}
a_2(T,\phi) = \frac{1}{U} - \int \frac{dk}{2\pi} \frac{1}{E_k}
+\int \frac{dk}{2\pi} \frac{1}{E_k}
\frac{\cos\phi + e^{-\beta E_k}}{\cosh(\beta E_k) + \cos\phi}.
\label{eq:a2_final}
\end{equation}
Since 
\begin{equation}
\frac{\cos\phi + e^{-x}}{\cosh x + \cos\phi}=1-\frac{\sinh x}{\cosh x + \cos\phi}
\end{equation}
and after inserting $-1/U_c$ and $+1/U_c$ we take finally
\begin{equation}
a_2(T,\phi) = \delta_u-\int_{\mathrm{BZ}} \frac{dk}{2\pi} \left[ 
\frac{\sinh(\beta E_k)}{E_k\big[\cosh(\beta E_k)+\cos\phi\big]}
- \frac{1}{|\epsilon_k|}
\right]
\end{equation}
where at $a_2=0$ we find the critical point where the thermal potential changes sign.

\end{document}